\documentclass[aps,prl,prabib,amsmath,amssymb,amsfonts,epsfig,smallcaptions,twocolumn,showpacs]{revtex4}
\usepackage{psfig}
\usepackage{graphicx}

\newcommand{\eqname}[1]{\label{eq:#1}}
\newcommand{\bgar}{\begin{eqnarray}}
\newcommand{\enar}[1]{\label{eq:#1}\end{eqnarray}}
\newcommand{\valass}[1]{\left|#1\right|}
\newcommand{\norme}[1]{\left\|#1\right\|}
\newcommand{\trace}[1]{\textrm{Tr}\left[#1\right]}
\newcommand{\ket}[1]{ | #1 \rangle }

\newcommand{\braket}[2]{ \langle #1 | #2 \rangle }

\newcommand{\ketbra}[2]{ | #1 \left\rangle \right\langle #2 |}

\newcommand{\mean}[1]{\overline{#1}}
\newcommand{\expect}[1]{\left\langle #1 \right\rangle}

\newcommand{\kk}{ {\bf k}}
\newcommand{\rr}{ {\bf r}}

\newcommand{\proj}{{\mathcal Q}}

\newcommand{\eq}[1]{(\ref{eq:#1})}
\newcommand{\al}[1]{^{(#1)}}
\newcommand{\Psihd}{\hat\Psi^\dagger}

\newcommand{\Psih}{\hat\Psi}

\newcommand{\Hamilt}{{\mathcal H}}
\newcommand{\ahd}{\hat a^\dagger}
\newcommand{\ah}{\hat a}

\begin{document}

\title{Condensate statistics in interacting Bose gases: exact results}

\affiliation{Laboratoire Kastler Brossel, \'Ecole Normale
Sup\'erieure, 24 rue Lhomond, 75231 Paris Cedex 05, France}

\author{Iacopo Carusotto}
\affiliation{Laboratoire Kastler Brossel, \'Ecole Normale
Sup\'erieure, 24 rue Lhomond, 75231 Paris Cedex 05, France}

\author{Yvan Castin}
\affiliation{Laboratoire Kastler Brossel, \'Ecole Normale
Sup\'erieure, 24 rue Lhomond, 75231 Paris Cedex 05, France}

\begin{abstract}
Recently, a Quantum Monte Carlo method alternative to the Path
Integral Monte Carlo method was developed for the numerical solution
of the $N$-boson problem; it is based on the
stochastic evolution of classical fields.
Here we apply it to obtain exact results for the occupation
statistics of the condensate mode in a weakly interacting trapped
one-dimensional Bose gas. 
The temperature is varied across the critical region down to temperatures
lower than the trap level spacing. We verify that the
number-conserving Bogoliubov theory gives accurate predictions
provided that the non-condensed fraction is small.
\end{abstract}



\date{\today}

\maketitle

The first achievement of Bose-Einstein condensation in weakly interacting 
atomic gases in 1995 has renewed the interest on basic aspects of the
Bose-Einstein condensation phase transition~\cite{stringari}.
In particular, an intense theoretical activity has been recently
devoted to the study of the occupation statistics of the
condensate mode: although no experimental result is available yet, we
expect that it would provide a
crucial test for many-body theories since, contrarily to more common
one-body observables like the density, it involves arbitrarily high
order correlation function of the quantum Bose field.
While there are now well established results for the ideal Bose
gas~\cite{IdealGas},
calculations for the weakly interacting Bose gas have
been performed within the framework of mean-field
approximation only~\cite{StringariN0,ScullyN0,Carthago}. The
intermediate regime around the critical temperature where the
mean-field theory fails has therefore been left unexplored.
Some controversy is still open about the validity of the Bogoliubov
approach even at temperatures much lower than the critical
temperature~\cite{Wilkens}. In the absence of experimental results, 
it is then very interesting to have exact theoretical results on the
statistics of condensate occupation.

There exists an exact analytical solution to the bosonic $N$ body
problem~\cite{Lieb}, but it is restricted to the spatially homogeneous
one-dimensional case.
From the side of numerics, the Quantum Monte Carlo method based on the
Path Integral Monte Carlo technique is in principle able to give
exact predictions for any observable of the gas~\cite{Ceperley} and was
successfully used to calculate the mean condensate
occupation~\cite{Krauth} and the critical temperature~\cite{Tc}.
In the Path Integral formulation, however, the position representation is
priviledged which makes the calculation of highly non-local observables like
the condensate occupation probabilities rather involved. 

In this paper, we use instead a recently developed 
Quantum Monte Carlo method based on the stochastic evolution of
classical fields~\cite{GPstoch,GPstochT}.
This new method has a much broader range of applicability than the
standard Path Integral Monte Carlo technique: it can be applied to
bosonic systems with complex wavefunctions, and even to interacting
Fermi systems~\cite{Chomaz}.
As compared to the Positive-$P$ method of quantum optics~\cite{PosP}, 
it has the decisive advantage of having been proven to be convergent.
In this paper, we perform the first non-trivial application of this
new method, calculating for the first time the exact distribution
function of the number of condensate particles in the presence of
interactions, for temperatures across the Bose-Einstein condensation
temperature.

An ultracold trapped interacting Bose gas in $D$ dimensions can be
modelled in a second-quantization formalism by the Hamiltonian
\begin{multline}
\label{eq:Hamilt}
{\mathcal H}=\sum_\kk \frac{\hbar^2 \kk^2}{2m} \ahd_\kk \ah_\kk +
\frac{g_0}{2} \sum_\rr dV\,\Psihd(\rr)\Psihd(\rr)\Psih(\rr)\Psih(\rr)\\
+\sum_\rr dV\,U_{\rm ext}(\rr) \Psihd(\rr)\Psih(\rr);
\end{multline}
the spatial coordinate $\rr$ runs on a discrete orthogonal lattice 
of ${\mathcal N}$ points with periodic boundary conditions; $V$ is the
total volume of the quantization box and $dV=V/{\mathcal N}$ is the
volume of the unit
cell of the lattice. $U_{\rm ext}(\rr)$ is the external trapping
potential, $m$ is the atomic mass and interactions are
modeled by a two-body discrete delta potential with a coupling
constant $g_0$.
The field operators $\Psih(\rr)$ satisfy the usual Bose commutation
relations $[\Psih(\rr),\Psihd(\rr')]=\delta_{\rr,\rr'}/dV$ and
can be expanded on plane waves according to $\Psih(\rr)=\sum_\kk
\ah_\kk e^{i\kk\rr}/\sqrt{V}$ with $\kk$ restricted to the
first Brillouin zone of the reciprocal lattice. 
In order for the discrete model
to correctly reproduce the underlying continuous field theory, the
grid spacing must be smaller than the macroscopic length scales
of the system like the thermal wavelength and the healing
length.

In this paper, we assume that the gas is at thermal equilibrium at 
temperature $T$ in the canonical ensemble so that the
unnormalized density operator is
${\rho}_{eq}(\beta)=e^{-\beta\Hamilt}$ with $\beta=1/k_B T$. 
It is a well-known fact of quantum statistical mechanics that this
density operator can be obtained as the result of an imaginary-time
evolution:
\begin{equation}\label{eq:ExactEvImT}
\frac{d{\rho}_{eq}(\tau)}{d\tau}=-\frac{1}{2}\left[{\mathcal
H}\,{\rho}_{eq}(\tau)+{\rho}_{eq}(\tau)\,{\mathcal H}\right].
\end{equation}
during a ``time'' interval $\tau=0\rightarrow \beta$ starting from the infinite
temperature  value ${\rho}_{eq}(\tau=0)=\textbf{1}_{\rm N}$.
We have recently shown~\cite{GPstochT}  that the solution of
the imaginary-time evolution \eq{ExactEvImT} can be written exactly as a
statistical average of Hartree operators of the form 
\begin{equation}\label{eq:Ansatz}
\sigma=\ketbra{N:\phi_1}{N:\phi_2}
\end{equation}
where, in both the bra and the ket, all $N$ atoms share the same (not
necessarily normalized) wave functions $\phi_\alpha$ ($\alpha=1,2$).
For the model Hamiltonian \eq{Hamilt} here considered, this holds if
each $\phi_\alpha$ evolves according to the Ito stochastic
differential equations:
\begin{multline}
d\phi_\alpha(\rr)=-\frac{d\tau}{2}\left[\frac{p^2}{2m}+U_{\rm ext}(\rr)+
g_0(N-1)\frac{\valass{\phi_\alpha(\rr)}^2}{\norme{\phi_\alpha}^2}\right.\\
-\left.\frac{g_0(N-1)}{2}\frac{\sum_{\rr'} dV\,\valass{\phi_\alpha(\rr')}^4}{\norme{\phi_\alpha}^4}\right]\phi_\alpha(\rr)+dB_\alpha(\rr)
\eqname{stoch}
\end{multline}
with a noise term given by
\begin{equation}
dB_\alpha(\rr)=i\sqrt{\frac{d\tau\, g_0}{2V}}
\proj_\alpha\big[\phi_\alpha(\rr)\sum_{\kk>0}(e^{i(\kk\cdot\rr+\theta_\alpha(\kk))}+\textrm{c.c.})\big]
\eqname{dB}\end{equation}
where the projector $\proj_\alpha$ projects orthogonally to
$\phi_\alpha$, the index $\kk$ is restricted to a half space and the
$\theta_\alpha(\kk)$ are independent random
angles uniformly distributed in $[0,2\pi]$.

Starting from this very simple but exact stochastic formulation,
a Monte Carlo code was written in order to
numerically solve the stochastic differential equation \eq{stoch}.
As a first step, a sampling of the infinite temperature density operator
 has to be performed in terms of a finite
number of random
wave functions $\phi\al{i}$. Since the effective
contributions of the different realizations to the final averages can be
enormously different, the statistics of the Monte Carlo results can be
improved by using an {\em importance sampling}
technique~\cite{Ceperley} so 
to avoid wasting computational time. 
This is done using the identities
\begin{equation}
\textbf{1}_{\rm N}=\int_1\!{\mathcal
D}\phi\,\ketbra{N:\phi}{N:\phi}=\int_1\!P_0[\phi]{\mathcal
D}\phi\,\frac{\ketbra{N:\phi}{N:\phi}}{P_0[\phi]}
\end{equation}
where the integration is performed over the unit sphere
$\norme{\phi}^2=\braket{\phi}{\phi}=\sum_\rr dV\,\valass{\phi}^2=1$
and where the {\em a priori} distribution function $P_0[\phi]$ can be freely
chosen in order to maximize the efficiency of the calculation. For the
numerical calculations here reported, the following $P_0[\phi]$ has
been used 
\begin{equation}
P_0[\phi]=\norme{e^{-h_{GP}\beta/2}\ket{\phi}}^{2N}
\eqname{P0}
\end{equation}
since it joined the possibility of a simple sampling with a fast convergence. 
In this expression, $h_{GP}$ is the Gross-Pitaevskii
Hamiltonian $h_{GP}=\frac{p^2}{2m}+U_{\rm
ext}(\rr)+N\,g_0\valass{\phi_0(\rr)}^2-\mu$, $\phi_0$ is the wave
function which minimises the Gross-Pitaevskii energy functional and
$\mu$ is the corresponding chemical potential. 
With respect to the ideal Bose gas distribution function
previously used~\cite{GPstochT}, the present form \eq{P0} for $P_0[\phi]$ 
has the advantage of taking into account the fact that the condensate
mode can be strongly modified by interactions. More details on the
actual sampling of $P_0[\phi]$ can be found in~\cite{GPstochT}.

Each realization $\phi\al{i}_{1,2}$ is then let evolve 
according to the stochastic evolution in imaginary-time \eq{stoch}
from its $\tau=0$ value $\phi\al{i}_{1,2}=\phi\al{i}$ to the
inverse temperature of interest $\tau=\beta$. 
The expectation values of any observable at temperature $T$ can
then be calculated as averages over the Monte Carlo realizations;
e.g., the partition function $\trace{\rho}$ is given by 
$\trace{\rho}=\mean{\braket{\phi_2}{\phi_1}^{N}}$.
In particular, the condensate wavefunction $\phi_{\rm BEC}(\rr)$ is
the eigenvector of the one-body density matrix
\begin{equation}
\eqname{OnePart}
\expect{\Psihd(\rr')\Psih(\rr)}=\frac{1}{\trace{\rho}}\; \mean{\phi_1(\rr)\phi_2^*(\rr')\braket{\phi_2}{\phi_1}^{N-1}}
\end{equation}
corresponding to the largest eigenvalue, that is to the largest mean 
occupation number~\cite{Onsager}.
The complete probability distribution $P(N_0)$ for the occupation statistics of
the condensate mode is obtained {\em via} the expression $P(N_0)={\mathbf
Tr}[{\hat {\mathcal P}}_{N_0}\rho]\;{\mathbf Tr}[\rho]^{-1}$ where ${\hat
{\mathcal P}}_{N_0}$ projects onto the subspace in which the condensate mode
contains exactly $N_0$ atoms; in terms of the ansatz \eq{Ansatz}, 
the expectation value of the projector can be written as
\begin{equation}
{\mathbf Tr}[{\hat {\mathcal P}}_{N_0}\rho]=\frac{N!}{N_0!\,(N-N_0)!}\;\mean{(c_2^{*}c_1)^{N_0}\;
\braket{\phi_2^\perp}{\phi_1^\perp}^{N-N_0}}
\eqname{OccStat}
\end{equation}
where we have split $\phi_{\alpha}(\rr)$ as $c_\alpha\phi_{\rm BEC}(\rr)+\phi_\alpha^\perp(\rr)$ 
with $\phi_\alpha^\perp(\rr)$ orthogonal to the condensate wavefunction.

\begin{figure}[htbp]
\psfig{figure=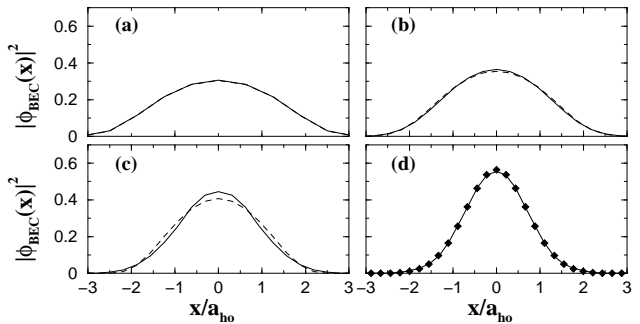,width=3.25in}
\caption{Solid lines: exact results for the condensate wavefunction
density profile
$|\phi_{\rm BEC}(x)|^2$
for $N=125$ atoms in a one-dimensional harmonic trap at
different temperatures $k_B T/\hbar \omega=1$ (a), $10$ (b), $20$ (c),
$50$ (d). $\omega$ is the trap frequency, $\valass{\phi_{\rm BEC}}^2$ is in units of the inverse 
of $a_{\rm ho}=\sqrt{\hbar/m\omega}$. 
The coupling constant is $g_0=0.08\,\hbar\omega\,a_{\rm
ho}$. Comparison with Bogoliubov predictions (dashed lines) in
(a,b,c) and with harmonic oscillator ground state (diamonds) in
(d). In (a) the two lines are indistinguishable on the scale of the
figure.
Monte Carlo calculations were performed using $3\cdot 10^4$ realisations on a
${\mathcal N}=16$ (a), $64$ (b,c) or $128$ point (d) grid.
\label{fig:foc}}
\end{figure}

\begin{figure}[htbp]
\psfig{figure=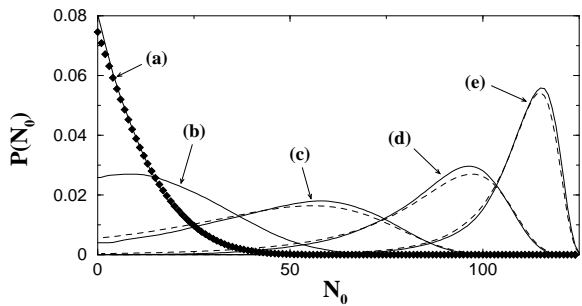,width=3.in}
\caption{Solid lines: exact results for the distribution function
$P(N_0)$ of the number of condensate atoms 
for decreasing temperatures from left to right: $k_B T/\hbar\omega=50$
(a), $33$ (b), $20$ (c), $10$ (d), $5$ (e). Same parameters as in
fig.\ref{fig:foc}.
For curves (c,d,e), comparison with Bogoliubov
predictions (dashed lines); for curve (a), comparison
with the ideal gas (diamonds). The Monte Carlo calculations
were performed using $3\cdot 10^4$ realisations on a grid of ${\mathcal N}=128$
(a,b) or ${\mathcal N}=64$ points (c,d,e).
\label{fig:stat_h}}
\end{figure}

\begin{figure}[htbp]
\psfig{figure=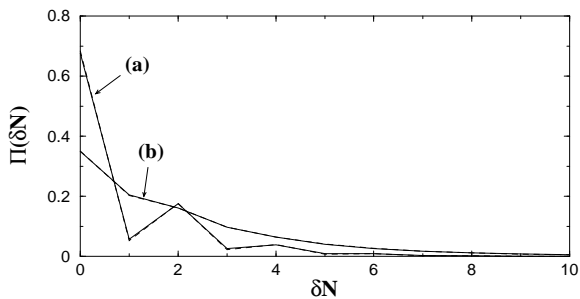,width=3.in}
\caption{Solid lines: exact results ($10^5$
realizations on a ${\mathcal N}=16$ point grid) for the 
distribution function $\Pi(\delta N)=P(N_0=N-\delta N)$ of the number of
non-condensed atoms at low temperatures $k_B T/\hbar
\omega=\,0.4$ (a) and $1$ (b). Same parameters as in fig.\ref{fig:foc} and
fig.\ref{fig:stat_h}. Dashed lines: Bogoliubov predictions. For $k_B
T/\hbar\omega = 0.4$, the two lines are indistinguishable on the scale of the
figure.
\label{fig:stat_l}}
\end{figure}

In fig.\ref{fig:foc}-\ref{fig:stat_l} we have summarized the results
of the Monte Carlo calculations for a one-dimensional, harmonically
trapped gas with repulsive interactions. 
In fig.\ref{fig:foc}, we have plotted the condensate wavefunction density profile 
$|\phi_{\rm BEC}(x)|^2$ for different values of the
temperature, see solid lines.
As the temperature decreases, the enhanced atomic density at the center of
the cloud gives a stronger repulsion among the condensate
atoms and therefore a wider condensate wavefunction. In this low
temperature regime, we have found a good
agreement between the Monte Carlo results and the Bogoliubov prediction for the
condensate wavefunction
including the first correction to the Gross-Pitaevskii equation due to
the presence of non-condensed atoms~\cite{BogoCastinDum}, see dashed
lines in fig.\ref{fig:foc}. This was expected, since the numerical
examples in this paper are in the weakly interacting regime
$n\xi\approx 15 \gg 1$, where $n$ is the density at the trap center and
$\xi=(\hbar^2/2mg_0n)^{1/2}$ is the corresponding healing length. 
For high temperatures, the wavefunction $\phi_{\rm BEC}(x)$ tends to
the harmonic oscillator ground state.

The wavefunction $\phi_{\rm BEC}(x)$ can then be used to determine the
occupation statistics of the condensate mode {\em via} \eq{OccStat}.
The signature of Bose condensation is apparent in
fig.\ref{fig:stat_h}: the occupation statistics of the condensate mode at
high temperatures has the same shape as a non-interacting, thermally
occupied mode, its maximum value being at $N_0=0$.
For decreasing $T$, $P(N_0)$ radically changes its shape; at low $T$, 
its maximum is at a non-vanishing value of $N_0$ and its shape is strongly asymmetric with a longer
tail going towards the lower $N_0$ values, as already noticed in the
mean-field approach of~\cite{Carthago}.
This transition ressembles the one occurring in a laser
cavity for a pumping rate which goes from below to above threshold~\cite{BECLaser}.

For even lower values of the temperature, see fig.\ref{fig:stat_l}, most of the atoms are in the
condensate and the probability distribution
$P(N_0)$ tends to concentrate around values of $N_0$ close to $N$. 
However, interactions prevent the atomic sample from being totally
Bose condensed even at extremely low temperatures.
Moreover, the probability of having an odd number of non-condensed
atoms is  strongly reduced with respect
to the probability of having an even number (see the strong
oscillations in the $k_B T=0.4\,\hbar\omega$ curve of fig.\ref{fig:stat_l}). A
quantitative interpretation of this effect in terms of the Bogoliubov
approximation will be given in the following part of the paper.

We now compare the exact results with the prediction of the number-conserving
Bogoliubov approximation~\cite{BogoGardiner,BogoCastinDum}.
The Bogoliubov Hamitonian has the form
\begin{equation}
\Hamilt_{\rm Bog}=\frac{1}{2}dV\,\left({\vec \Lambda}^\dagger,{\vec
\Lambda} \right)\cdot\eta{\mathcal L}\left(\begin{array}{c}{\vec
\Lambda} \\ {\vec \Lambda}^\dagger \end{array}\right)
\eqname{BogHam}
\end{equation}
with
\begin{equation}
{\mathcal L}=\left(\begin{array}{cc} h_{\rm GP}+\proj_0
Ng\valass{\phi_0}^2\proj_0 & \proj_0
Ng\phi_0^2\proj_0^* \\ -\proj_0^*
Ng\phi_0^{*2}\proj_0 & -h_{\rm GP}-\proj_0^*
Ng\valass{\phi_0}^2\proj_0^*\end{array} \right).
\end{equation}
The matrix $\eta$ is defined according to
\begin{equation}
\eta=\left(\begin{array}{rr} {\bf 1} & 0 \\ 0 & -{\bf 1}  \end{array}
\right)
\end{equation}
and the projector $\proj_0$ projects orthogonally to the
Gross-Pitaevskii ground state $\phi_0$.
The operators ${\vec \Lambda}$ are defined according to $({\vec
\Lambda})_\rr={\hat A_0^\dagger}\Psih_\perp(\rr)$ where $\Psih_\perp(\rr)$ is the projection of the 
field operator orthogonally to $\phi_0$ and where ${\hat
A_0}\ket{N_0\!:\!\phi_0}=\ket{N_0-1\!:\!\phi_0}$ if $N_0>0$ and $0$ otherwise.
Neglecting the possibility of completely emptying the condensate mode, ${\hat
A_0}$ and ${\hat A_0^\dagger}$ can be shown to commute. Under this
approximation, the ${\vec \Lambda}$ are bosonic operators.
The probability $\Pi(\delta N)$ of having $\delta N$ non-condensed atoms
can be calculated within the Bogoliubov approximation 
by means of the characteristic function~\cite{ScullyN0}
\begin{equation}
f(\theta)=\sum_{\delta N} \Pi(\delta N)\,
e^{i\,{\delta N}\,\theta}\simeq\trace{e^{i\theta\,{\delta {\hat N}}}\frac{1}{{\mathcal
Z}}e^{-\beta \Hamilt_{\rm Bog}}}
\end{equation}
with $\delta {\hat N}=dV\,{\vec \Lambda}^\dagger\cdot{\vec \Lambda}$
and ${\mathcal Z}=\trace{e^{-\beta \Hamilt_{\rm Bog}}}$; 
when the probability of emptying the condensate mode is not totally negligible, 
the Bogoliubov prediction for $\Pi(\delta N)$ has to be truncated to the physically
relevant $\delta N\leq N$ values and then renormalized to $1$
.
The details of the calculation will be given elsewhere, we go here to the result
\begin{equation}
f(\theta)\simeq\prod_{j=1}^{2({\mathcal N}-1)}
\left[\frac{2\lambda_j}{1+\lambda_j-e^{i\theta}(1-\lambda_j)}\right]^{1/2}
\end{equation}
in which the $\lambda_j$ are the eigenvalues of
${\mathcal M}=\eta\tanh(\beta{\mathcal L}/2)$; from such an expression, the
occupation probabilities $\Pi(\delta N)$ are immediately determined by
means of an inverse Fourier transform. The results are shown as dashed lines in
figs.\ref{fig:stat_h} and \ref{fig:stat_l} 
where they are compared to the
exact results from Monte Carlo simulations; the agreement is
excellent provided the non-condensed fraction is small, i.e. at
sufficiently low temperatures. 


In order to understand the oscillating behaviour shown by the lowest
temperature curve of fig.\ref{fig:stat_l}, we calculate the ratio
of the probability $\Pi_{\rm odd}$ of having an odd value of $\delta N$ to the
probability $\Pi_{\rm even}$ of having an even value:
\begin{equation}
\frac{\Pi_{\rm odd}}{\Pi_{\rm
even}}=\frac{f(0)-f(\pi)}{f(0)+f(\pi)}\simeq\frac{1-R}{1+R}
\eqname{R}
\end{equation}
where $R=\prod_j \lambda_j^{1/2}=\det[{\mathcal M}]^{1/2}=
\prod_m \tanh (\beta\epsilon_m/2)$ and $m$ runs over the
${\mathcal N}-1$ eigenenergies $\epsilon_m$ of the Bogoliubov
spectrum. 
In order to have oscillations in $\Pi(\delta N)$, the ratio \eq{R} must be  small as compared to $1$ which requires a 
temperature smaller than the energy of the lowest Bogoliubov mode.
These oscillations are therefore a property of the ground state of the system.
In the Bogoliubov approximation, the Hamiltonian is quadratic in the field operators
so that its ground state is a squeezed vacuum, which indeed contains only even values of $\delta N$,
a well known fact of quantum optics~\cite{WallsMilburn}. From a condensed matter
point of view, several existing variational ansatz for the ground
state corresponding to a condensation of pairs also exhibit this
property~\cite{Girardeau,Leggett}.
Physically, this is ultimately related to the fact that the leading
interaction process changing the condensate occupation at zero temperature, that is the scattering of two
condensate particles into two non-condensed modes and {\em vice-versa}, does
not change the parity of $\delta N$, as it is apparent in the
Bogoliubov Hamiltonian \eq{BogHam}.

In conclusion, we have determined with a newly developed stochastic
field Quantum Monte Carlo
method the exact distribution function of the number $N_0$ of
condensate particles
in a one-dimensional weakly interacting trapped gas. The signature of
Bose condensation is the appearance of a finite value for
the most probable value of $N_0$.
At temperatures below the trap oscillation frequency, configurations with
an odd number of non-condensed particles are strongly suppressed,
which we interpret successfully within the Bogoliubov approximation.
Possible extensions of this work are (i) the determination of
critical temperature shift for an interacting Bose gas
in the limit of a vanishing scattering length, still subject 
of some controversy \cite{controversy}, (ii)
exact calculations for finite 
temperature Bose gases with vortices in rotating traps,  and (iii) the 
determination of the BCS critical temperature in two-component Fermi gases.

\begin{acknowledgments}
This work was stimulated by interactions with I. Cirac. We acknowledge useful 
discussions with A. Leggett and J. Dalibard.
I.C. acknowledges a Marie Curie grant from the EU under contract number
HPMF-CT-2000-00901.
Laboratoire Kastler Brossel is a Unit\'e de
Recherche de l'ENS et de l'Universit\'e Paris
6, associ\'ee au CNRS.
\end{acknowledgments}

\end{document}